**From Amorphous to Amorphous-Crystalline Mixed-Phase Boron Nitride: Evolution of the Thermal and Elastic Properties**

Jiaqi Yang[1,2]*, Thomas Souvignet[3]*, Onurcan Kaya[1]*, Peng Xiao[1], Emigdio Chavez-Angel[1], Daniel Capolat Palomar[1,2], Zuzanna Ewa Kaczmarska[4], Bartlomiej Graczykowski[4], Javier Rodriguez Viejo[1,2], Clivia M. Sotomayor Torres[5], Catherine Marichy[3], Catherine Journet[3], Stephan Roche[1,6], Marianna Sledzinska[1]

[1] Catalan Institute of Nanoscience and Nanotechnology (ICN2), CSIC and BIST, Campus UAB, Bellaterra, Barcelona 08193, Spain

[2] Faculty of Physics, Autonomous University of Barcelona, Bellaterra, 08193 Barcelona, Spain

[3]Universite Claude Bernard Lyon 1, CNRS, LMI, UMR 5615, Villeurbanne F-69100, France

[4] Faculty of Physics and Astronomy, Adam Mickiewicz University, Uniwersytetu Poznańskiego 2, Poznan 61-614, Poland

[5]International Iberian Nanotechnology Laboratory, Av. Mestre José Veiga s/n 4715-330 Braga, Portugal

[6] ICREA, Passeig Lluis Companys 23, 08010 Barcelona, Spain.

*These authors contributed equally

Corresponding authors: marianna.sledzinska@icn2.cat, stephan.roche@icn2.cat, catherine.journet@univ-lyon1.fr

**Abstract**

Amorphous boron nitride (a-BN) is a promising dielectric and protective coating, yet its nanoscale heat dissipation and elastic response remain poorly quantified. Here we synthesize a variety of BN thin films by borazine-based chemical vapor deposition (800–1000 °C) and study the temperature-driven structural transition from fully amorphous networks to mixed amorphous–crystalline films with embedded BN nanocrystallites. Frequency-domain thermoreflectance data show an ultralow, thickness-dependent cross-plane thermal conductivity for a-BN ($k_{out}$ < 0.5 W m$^{-1}$ K$^{-1}$ for 10–40 nm), which increases systematically with crystalline order up to ~1.5 W m$^{-1}$ K$^{-1}$. Micro-Brillouin light scattering and finite-element modelling reveal a concomitant stiffening, with Young's modulus rising from 7.5 ± 0.7 GPa (800 °C) to 53 ± 5 GPa (1000 °C). Green–Kubo molecular dynamics simulations rationalize these trends via bonding topology and vibrational transport, and highlight how oxygen, hydrogen and carbon impurities and composition provide practical knobs to further tune the thermal and mechanical responses in BN films for improving nano-electronics, interconnects and coating applications.

2D Boron nitride (BN) has gained significant attention as an electrical insulator. Among its different polymorphs, hexagonal BN (h-BN) shows remarkable electrical, optical and thermal properties. It has been commonly used as an efficient dielectric material for 2D material-based transistors, given its ultra-flat surface and dielectric constant of ~3 in the out-of-plane direction [1–3]. On the other hand, amorphous BN (a-BN) is an emerging material with low dielectric constant ((1.78-1.16 at 100 kHz and 1 MHz respectively), and high mechanical robustness, offering plenty of potential applications in electronics[4,5], batteries[6] and coatings[7]. Several a-BN synthesis approaches have been recently reported, such as pulsed laser deposition (PLD)[8], atomic layer deposition (ALD)[9,10] or chemical vapor deposition (CVD)[11]. Among these techniques CVD enables uniform large-scale and low-temperature deposition, which can be a viable alternative for conventional dielectric oxides ($Al_2O_3$, $HfO_2$) for high-performance electronics or interconnect technology.

The reliability of the electronic devices is essential and directly linked to the thermal and mechanical properties of the films used for their fabrication. A crucial location for the formation of hotspots usually lies at the interface between the dielectric layer and metallic gate contacts. As far as thermal properties, standard thin oxide gate dielectrics, including $SiO_2$, $Al_2O_3$, $HfO_2$ show low thermal conductivities $k$ (in the order of 1W/mK), which can be harmful for the overall device thermal management. On the other hand h-BN exhibits remarkable thermal properties, with thermal conductivity reaching 500 W/mK and 5-10 W/mK, in the in-plane and cross-plane directions, respectively [12–14], enabling an efficient thermal management of 2D electronic devices [15,16]. Amorphous materials have in general significantly lower thermal conductivity, but to date the properties of a-BN remain largely unexplored, in terms of thickness dependence and influence of typical doping elements, such as oxygen and carbon. Moreover, changes in structural properties, such as the $sp^2/sp^3$ bonding fraction, may affect thermal transport in a-BN by modifying the local bonding network and the associated vibrational scattering pathways. The low thermal conductivity is intrinsically related to low Young modulus of such materials, resulting in flexibility and resistance to brittleness, which makes them ideal for coatings.

Here, we report on structural, thermal and elastic characterization of a BN films prepared by CVD using borazine precursor as a function of deposition temperature (800 – 1000 ºC). The cross-plane $k$ of the aBN films was experimentally studied using frequency-domain thermoreflectance (FDTR) technique while the elastic properties were measured using Brillouin light scattering. The experimental findings were compared to the theoretical simulations, modelled via classical equilibrium molecular dynamic (EMD) approach. Using EMD calculations, thermal conductivity, and spatial features of lattice vibrations were explored. The effects of the degree of amorphization and local bonding configuration, including the sp2/sp3 ratio, on thermal conductivity were also examined. Further, we also probed the change of thermal properties against the typical contaminants, such as hydrogen and oxygen.

The CVD-grown BN thin films (*Materials and Methods* and Figure S1 and S2) with thickness in the range of a few tens of nanometers are uniform, as demonstrated in the SEM images in Figure S3. The crystallinity of the films is controlled by varying the growth temperature ranging from 800 to 1000 ºC (Figure 1a). Characteristic features of $sp^2$ B-N bonding are visible in the FTIR spectra (Figure 1b) at 1400-1300 $cm^{-1}$ and 820-780 $cm^{-1}$. Both bands sharpen and intensify as the CVD temperature increases, which is a sign of an improvement of the BN ordering. The Raman spectra (Figure 1c) confirm the presence of a band around 1370 $cm^{-1}$ in the film grown at 900 °C, assigned to crystalline $sp^2$ BN. The film grown at 1000 °C presents a band with a full width at half maximum above 40 $cm^{-1}$. This evidences an improvement in the crystallinity as a function of the CVD temperature. The fully amorphous phase (a-BN) is achieved for growth temperature of 800 °C, while partial crystallization is obtained as the temperature increases to 900 °C, resulting in a mixed amorphous-crystalline phase, where BN crystallites are embedded in an amorphous matrix. Finally, above 1000 °C, the film becomes predominantly polycrystalline.

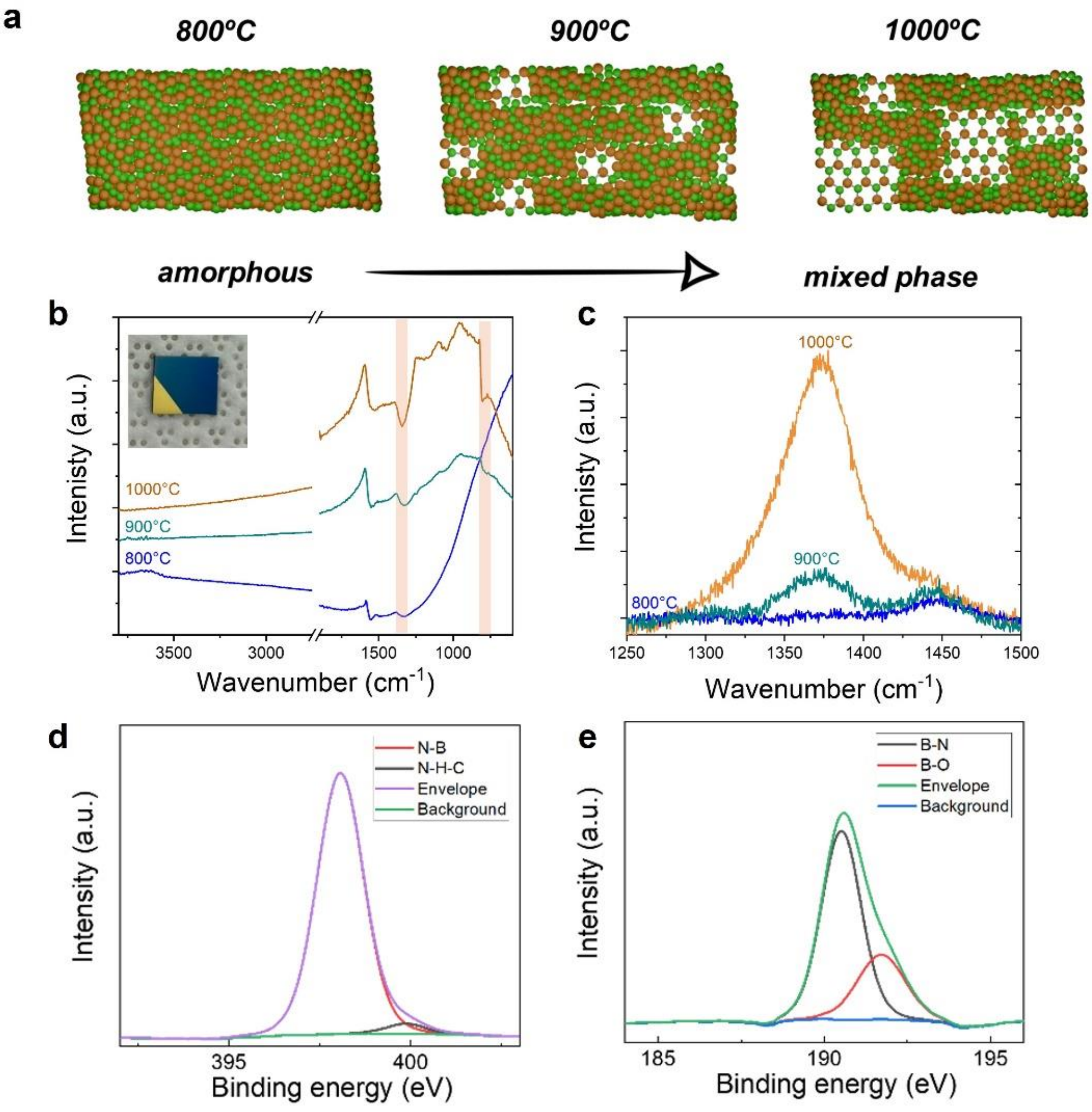

**Figure 1. Structural characterization of BN films.** a) Schematics of the transition from the amorphous to the mixed phase structure as a function of synthesis temperature. b) FTIR and c) Raman spectra recorded on CVD films grown at 800 °C, 900 °C and 1000 °C. The regions marked in orange correspond to $sp^2$ characteristic bands of BN. Inset: optical image of 1x1 cm silicon chip fully covered with a-BN film with Au transducer for FDTR measurements. Deconvoluted XPS d) nitrogen 1s and e) boron 1s spectra for amorphous BN sample grown at 800 ºC.

Figures 1 d,e and Figure S2 show the evolution of the boron and nitrogen bonds as a function of growth temperature. For all the samples a clear presence of B-N bonds is seen, with a small presence of N-H-C and N-C bonds, indicating the presence of carbon contamination arising from air exposure. For the 800 ºC samples, B-O bonds are also present, which stem either from the growth process or the interaction of the sample with the ambient atmosphere. Indeed, partial oxidation of the film may occur due to highly reactive B-H and N-H bonds remaining in the amorphous structure. For higher synthesis temperatures the B-B bonds start to appear, and the corresponding XPS signal increases. The typical B/N ratio in the films is evaluated between 1.2 and 1.4, showing a slight boron excess. The summary of film thickness and elemental composition extracted from ellipsometry and XPS, respectively, is shown in Table S1.

To better understand the crystalline structure and nature of atomic bonding we conducted high resolution transmission electron microscopy (HRTEM) together with energy dispersive X-ray spectroscopy (EDS) and electron energy loss spectroscopy (EELS). A thin lamella was prepared from the a-BN synthesized at 800 ºC on top of Si substrate with Au protection layer. As shown in Fig. 2a, the sample exhibits fully amorphous structure. Indeed, no lattice fringes and no rings are visible in the selected area electron diffraction pattern (SAED), characteristic of amorphous materials. On the other hand, the BN synthesized at 1000 ºC and transferred to a TEM grid shows clear lattice fringes with spacing typical of $sp^2$ BN and a SAED pattern with clearly defined annular rings, which are characteristic of polycrystalline structure (Figure 2b). These observations further corroborate the conclusions derived from Raman spectroscopy analysis (Figure 1).

To correlate the structures observed by HRTEM with bonding configurations, we additionally compared EELS on the amorphous samples, with mechanically exfoliated h-BN flakes. As shown in Figure 2c, the EELS spectrum of h-BN exhibits typical near-edge fine structure: sharp $\pi^*$ resonance peaks approximately 192 eV (B k-edge) and 405 eV (N k-edge), followed by distinct $\sigma^*$ characteristic peaks at higher energy losses. In contrast, aBN exhibits significantly broadened $\pi^*$ and $\sigma^*$ peaks in the B k-edge region, occurring near 194 eV and 201.5 eV respectively, with broadly comparable intensities (the $\sigma^*$ peak being approximately 10% stronger). Interestingly, no signal from the $\pi^*$ or $\sigma^*$ peaks was detected in the N k-edge region of a-BN. However, the EDS spectrum confirms the presence of boron and nitrogen throughout the film (Figure S4). We propose this may represent a key characteristic of amorphization, wherein $\pi^*$ and $\sigma^*$ resonances are strongly suppressed at the N K-edge[17]. This suppression stems from the sample's lack of long-range, or even short-range, $sp^2$ coordination order. An extended computational analysis and discussion of the of $sp^2$ hybridization in both aBN and hBN can be found in the SI (Figures S5-S6).

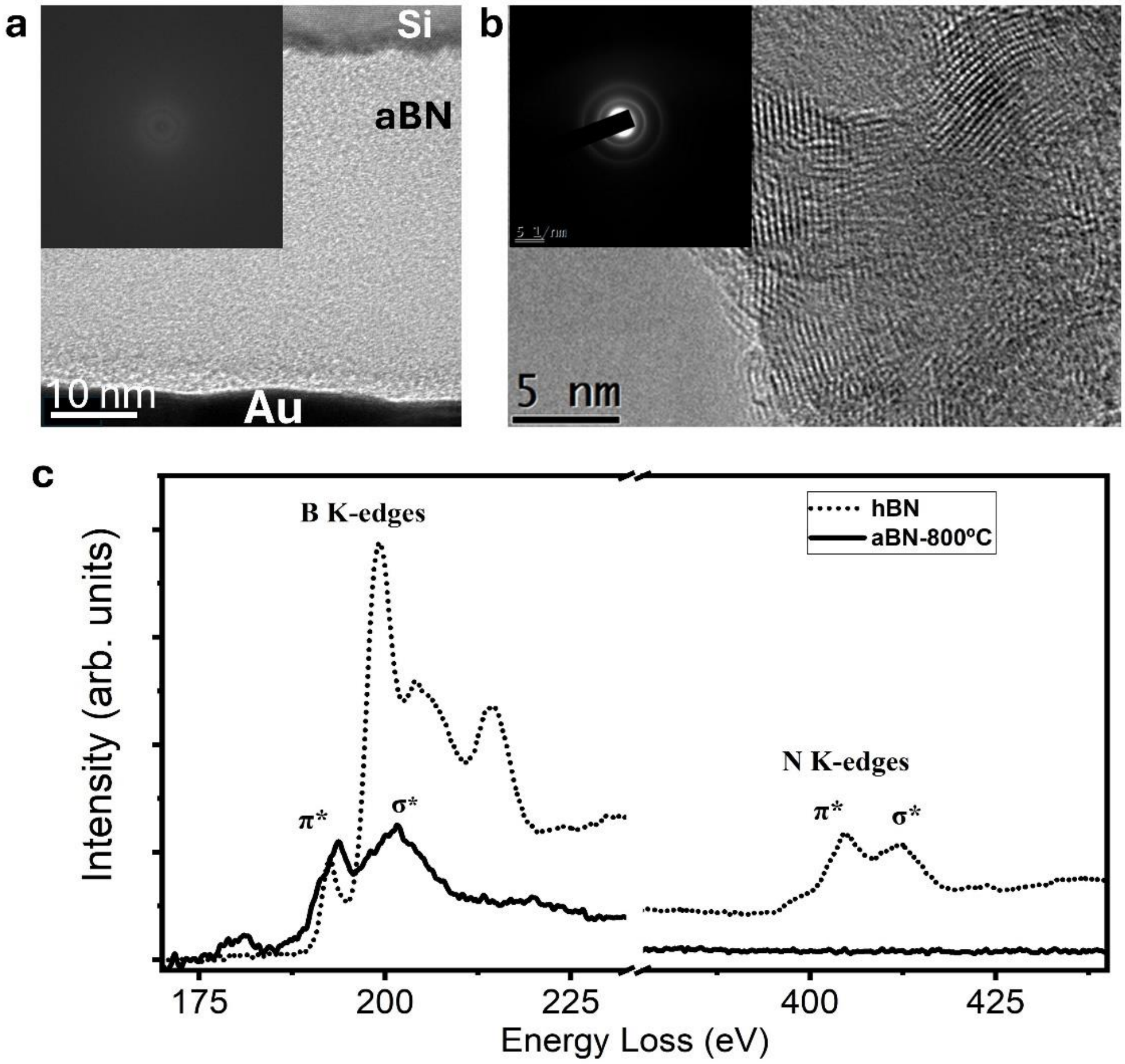

**Figure 2. High-resolution transmission electron microscopy** (a) Cross-sectional HRTEM image of BN-800 ℃ sample; (b) HRTEM image of BN-1000℃. Insets: SAED patterns. (c) EELS showing the peaks corresponding to boron and nitrogen K-edges of BN-800 ℃ and mechanically exfoliated h-BN reference sample.

Critical hotspots in electronic devices typically occur at the interface between the gate electrode and the dielectric material, due to high electric fields and high switching velocities. Here we employ frequency domain thermoreflectance (FDTR) technique[18–20] to study the metal-dielectric-silicon interfaces, as shown in Figure 3a (inset). Thermal properties, such as $k$, heat capacity $c_p$, or thermal interface conductance $G$ of the sample can be obtained by fitting an analytical model to the measured phase lag across the range of modulation frequencies (see Figures S7 and Table S2 for details). Cross-plane thermal conductivity $k_{out}$ was extracted using an analytical model to fit the data using a three-layer model over the frequency range given by the sensitivity analysis (Figure S8)[21,22]. Interfacial thermal conductances $G_{12}$and $G_{23}$ were set to conservative literature-based values of 20 and 30 MW/(m$^2$ k), respectively[23,24]. An example of experimental data together with an analytical fit for aBN grown at 800 ºC is shown in Figure 3a.

The thermal conductivity $k_{out}$ as a function of thickness and growth temperature is plotted in Figure 3b. The a-BN synthesized at 800ºC exhibits the lowest $k_{out}$, measured to be below 0.5 W/mK. The $k_{out}$ shows a slight increase (0.1 to 0.3 W/mK) in the thickness range between 10 and 40 nm. The ultralow thermal conductivity of aBN is attributed to the lack of crystalline structure and suppressed contribution of extended vibrational modes, propagons and diffusons. Similar examples of strong reduction of thermal conductivities of van der Waals materials via structure modification have been reported previously and include reduction of *k* of disordered $WSe_2$ to 0.05 W/mK, a factor of 6 smaller than the predicted minimum thermal conductivity [25] or $MoS_2$ to below 0.3 W/mK[26,27]. With the increasing synthesis temperature (BN-900ºC and BN-1000ºC) and improved crystalline structure, $k_{out}$ increases to up to 1.5 W/mK for the thickest samples. We also note that $k_{out}$ shows a systematic increase with the film thickness, for both amorphous and mixed-phase samples. For comparison, we studied mechanically exfoliated h-BN, with thickness ranging from 30 nm to 100 nm. While h-BN is an excellent thermal conductor in the in-plane direction (300-400 W/mK), it shows much lower $k_{out}$, due to the weak van der Waals bonding between the layers (Figures S9-S11). For the flakes with thickness of few tens of nm the $k_{out}$ ranges between 1 and 2 W/mK.

According to Debye's theory thermal transport in solid materials strongly depends on the phonon mean free path (MFP). The contributions from phonons with long MFPs are suppressed in a film of finite thickness, leading to lower thermal conductivity. A recent study found that phonons with MFP >100 nm at room temperature are responsible for most of the thermal transport in h-BN[28] . These results are in line with our measurements, which show an increase in $k_{out}$ from 1.2 to 3.5 W/mK for thickness from 35 to 100 nm (Figure S10). On the other hand, for polycrystalline 2D materials, grain boundary scattering has been proposed as a limiting factor for the maximum possible phonon MFPs. For the case of the mixed phase BN a clear increase in $k_{out}$ is observed, suggesting MFPs up to few tens of nanometers. Surprisingly, a thickness dependence of $k_{out}$ is also observed for the a-BN sample. We note that in other amorphous films, such as $SiO_2$ and $Al_2O_3$, such size dependence is not as remarkable[29]. This result underscores the significance of the contributions from propagons and diffusons to the thermal conductivity of amorphous materials, which can vary for each material.

If we further benchmark a-BN against other amorphous dielectric materials, such as $TiO_2$ , $Al_2O_3$ and $SiO_2$ ($k_{out}$ of 1, 1.3 W/mK [22] and 1.3 W/mK[30], respectively, for $t$ = 50 nm) we note that their thermal conductivities are of the same order of magnitude as BN studied in this work. This result positions the a-BN as a promising candidate for modern electronics. Furthermore, in amorphous materials, the atomic network topology and vibrational mode localization can lead to significant changes in thermal properties[31]. The presence of oxygen and hydrogen in our samples is also expected to contribute to their thermal properties, since the insertion of impurities not only changes the physical properties of the a-BN but also introduces chemical complexities to the system. In addition to atomic mass mismatch, disruptions in the bond structure or atomic network should significantly impede the propagation of vibrational energy. As found in our simulations, the variations in density and bonding type are directly associated with thermal transport properties, as explained in details in further sections [32].

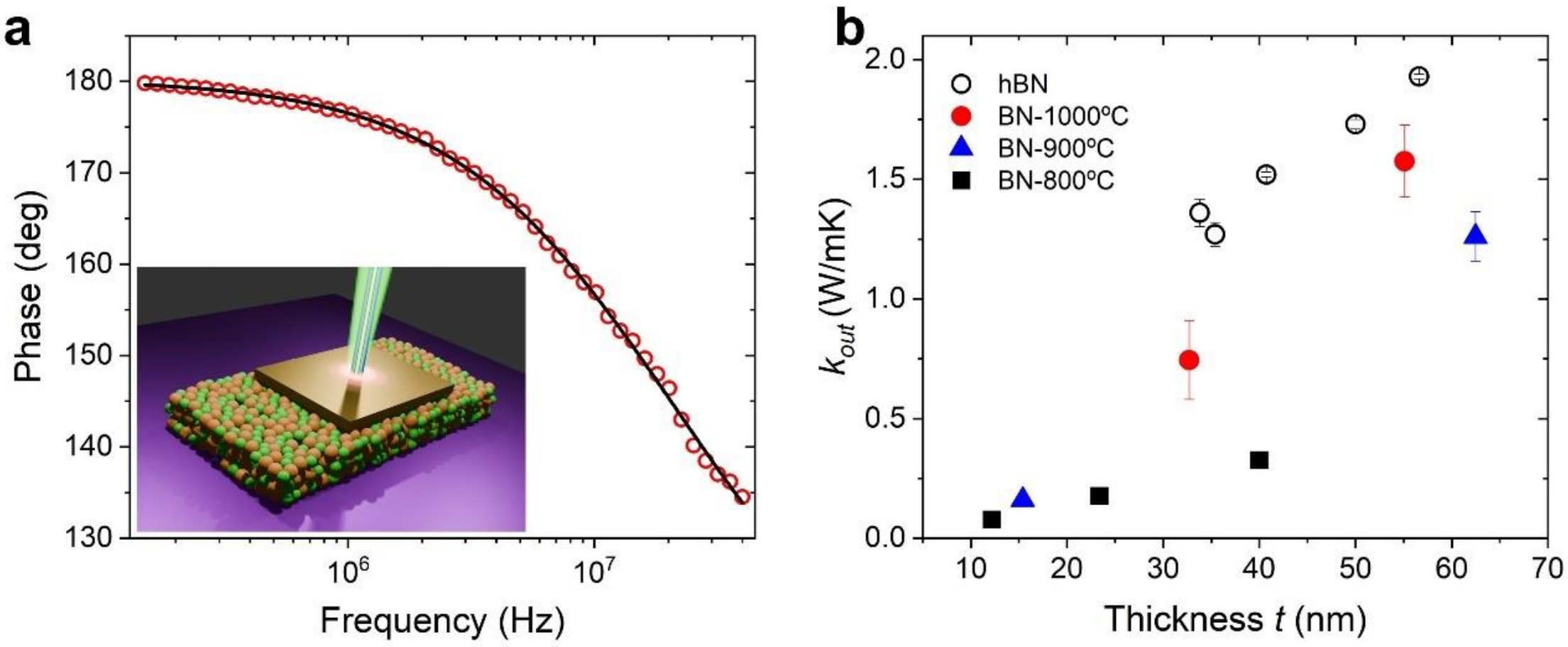

**Figure 3. Cross-plane thermal transport measurements** (a) experimental (circle) and fitted (line) phase lag signal as a function of modulation frequency measured for a-BN synthesized at 800ºC. Inset: Schematic of the FDTR measurements setup. (b) The cross-plane thermal conductivity of the CVD-grown BN films, compared to mechanically exfoliated h-BN. The error bars include both the experimental noise and uncertainties from various input parameters for the fitting.

To determine the elastic properties of the BN films grown at 800 °C and 1000 °C, the micro-Brillouin light scattering (μ-BLS) spectroscopy operating in the backscattering geometry was employed. The representative spectra shown in Figure 4a display peaks corresponding to the Rayleigh surface wave (RSW) and the Sezawa acoustic surface wave (SAW). The broader RSW peak observed for the BN film grown at 800 °C, compared with the film grown at 1000 °C, indicates shorter phonon lifetimes and greater structural inhomogeneity or surface roughness of the former sample. RSW dispersion relations reveal a notable difference in the elastic properties of the investigated BN films (Figure 4b). The FEM calculations performed in COMSOL Multiphysics yielded Young modulus values of 7.5 ± 0.7 GPa for BN-800˚C, and 53 ± 5 GPa for the BN-1000˚C. The significant increase in Young modulus with increasing synthesis temperature confirms the correlation between thermal treatment and the structural ordering of the BN films. Synthesis at higher temperatures promotes grain growth and reduces structural defects, such as vacancies and voids, thereby improving crystallinity and mechanical properties. In contrast, the lower modulus determined for the amorphous BN-800°C film is related to disorder and possible presence of nanopores, which contribute to enhanced phonon scattering and broader BLS peaks. We additionally performed measurements on h-BN flake transferred onto a silicon substrate. FEM fitting to the experimental data (Figure 4c) allowed determination of the elastic constants $c_{44}$ = 6.0 ± 1.0 GPa and $c_{33}$ = 24.5 ± 0.5 GPa. In the case of h-BN, the elastic constant $c_{13}$ = 0 GPa[33]; therefore, $c_{33}$ is equal to the out-of-plane Young modulus, $E_3$= $c_{33}$ (Figure 4d). The obtained $E_3$ value is in agreement with previous reports (27 GPa)[33], however, it is smaller than that of BN-1000˚C. Single-crystalline h-BN is a highly anisotropic material, and the in-plane Young modulus $E_1$ was previously reported to be 775 GPa (Figure 4d). Thus, in polycrystalline films, averaging effects arising from different crystallite orientations and sizes are expected[34]. Similarly, for thin polycrystalline BN films with turbostratic microstructure prepared using ALD, the elastic modulus was found to be between 100 and 150 GPa, depending on the annealing temperature[10]. In our case, low Young modulus of a-BN results in higher flexibility and resistance to brittleness, which makes it ideal for coating applications.

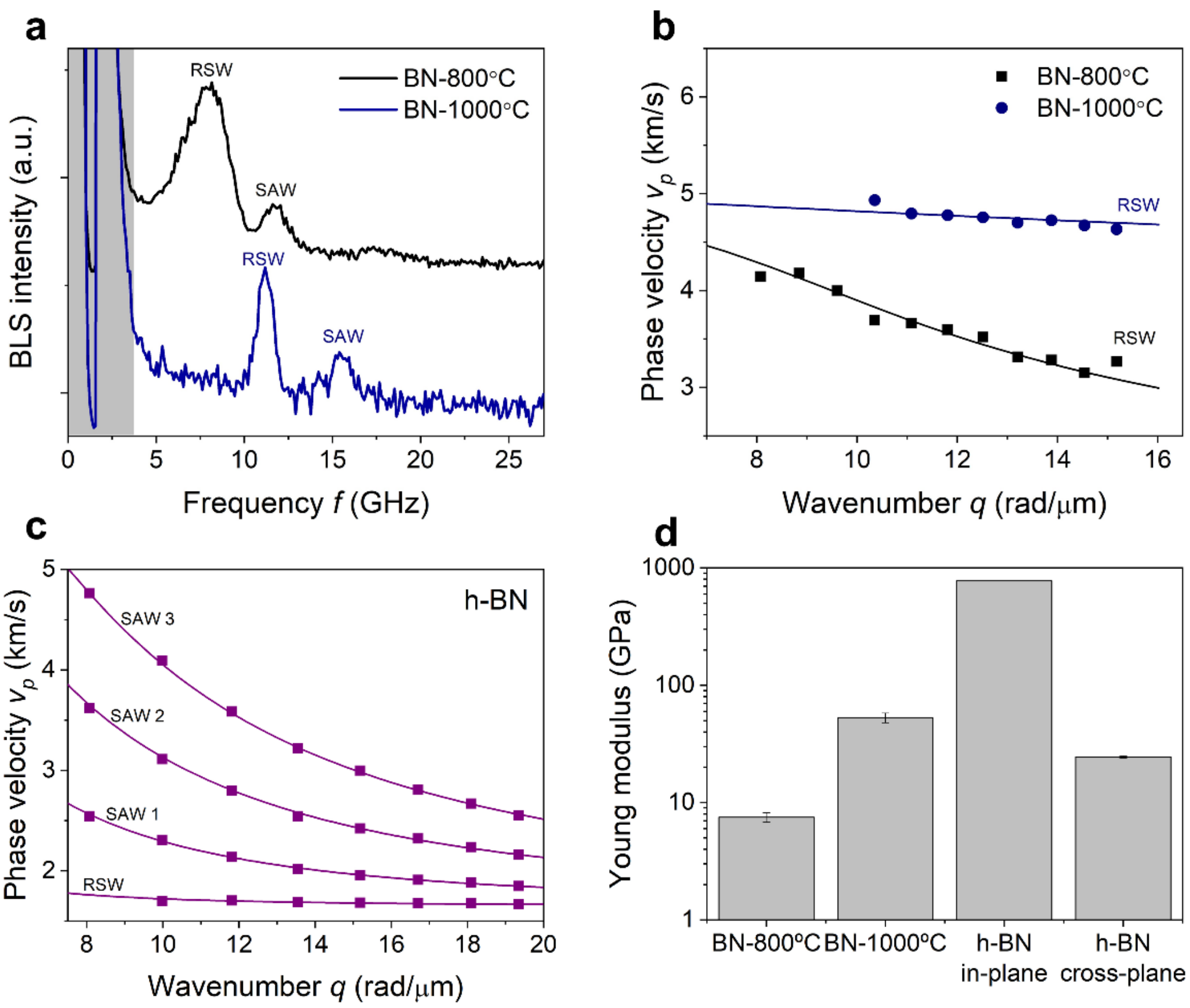

**Figure 4**. **Elastic properties of BN films grown at 800˚C and 1000˚C** (a) Representative µ-BLS spectra of the samples measured at a scattering angle of 40˚, corresponding to a wavenumber $q$ =15.18 rad/µm. The shaded gray area denotes the frequency range of elastic scattering. (b) Dispersion relations of the Rayleigh surface wave (RSW). (c) Dispersion relations of RSW and Sezawa acoustic surface waves (SAWs) in single-crystalline hexagonal BN flake (949 ± 2 nm in thickness) transferred onto a silicon substrate. Solid lines in (b) and (c) represent fits obtained from FEM calculations in COMSOL Multiphysics. (d) Comparison of the determined Young modulus values. The value of the in-plane Young modulus $E_1$ for h-BN is taken from[33], and the value of out-of-plane Young modulus $E_3$ is determined from the modeling shown in (c).

Finally, Green-Kubo molecular dynamics simulations were performed to further elaborate on the trends in lattice thermal conductivity as a function of morphology and composition for the case of a-BN. All structures were prepared using the melt-quench protocol described in *Materials and Methods* with a cooling rate of 1 K/ps (Figure 5a Inset). The effects of non-stoichiometry and contamination were studied separately to isolate their respective contributions to thermal transport.

The effect of non-stoichiometry was studied by varying the B/N ratio from 1.0 to 1.5. With increasing boron concentration, the excess boron is accommodated through the formation of B-B bonds and promotes the B-clusters within the structure. Moreover, with the increasing B/N ratio, $sp^3$ fraction is reduced significantly while $sp^1$ sites become more prevalent (Figure 5a and Figure S12-13), and density decreases from 2.21 to 1.93 g/cm$^3$ with accompanying nanovoid formation. These structural changes alter the vibrational density of states considerably, as shown in Figure 5b. The longer B–B bonds shift vibrational modes to lower frequencies and broaden the vibrational spectrum. With increasing B/N ratio, the low-frequency modes become more populated, and the vibrational density of states flattens. Moreover, higher-frequency modes above 50 THz become more prevalent and are mostly localized as shown by the participation ratio (Figure S14-15).

Thermal conductivity decreases monotonically from 1.43 W/mK at B/N = 1.0 to 1.09 W/mK at 1.5. The participation ratio also confirms that the non-stoichiometric samples carry a higher population of localized modes at low frequencies. Beyond this, increasing porosity, with density dropping from 2.21 to 1.93 g/cm$^3$, adds void boundaries that scatter the modes that do propagate through the network. The experimental XPS data show a B/N ratio between 1.2 and 1.4 in the CVD-grown films (Table S1), which according to the simulations produces a 15–25% reduction in thermal conductivity relative to stoichiometric a-BN before contamination effects are considered.

We next observe that the incorporation of carbon atoms produces a non-monotonic response in thermal conductivity, with $k$ increasing to 1.67 W/mK at 15% before declining at higher concentrations (Figure 5d). At low concentrations, carbon disperses through the B-N matrix and replaces homonuclear B-B and N-N bonds with heteronuclear B-C and C-N bonds. Since the atomic mass of carbon lies between boron and nitrogen, these bonds have vibrational frequencies closer to B-N than the homonuclear bonds they replace, reducing frequency mismatch. Carbon also reduces the population of undercoordinated sp-hybridized sites, where the lack of bonding constraints localizes vibrational modes and suppresses thermal transport. The increased $sp^3$ crosslinking raises density and reduces void volume (Figure S17), further suppressing phonon scattering at void boundaries.

Unlike carbon, hydrogen and oxygen both reduce thermal conductivity at concentrations above 5%, though distinct mechanisms. Hydrogen shows a slight improvement at 5% ($k$ = 1.50 W/mK) due to passivation of undercoordinated sites, which reduces the number of localized vibrational modes. At higher concentrations, the light mass of hydrogen introduces strong frequency mismatch with the B-N network, as B-H and N-H stretching modes appear above 50 THz, spectrally separated from the lower-frequency B-N vibrations that dominate thermal transport. Beyond 5%, hydrogen progressively lowers coordination and opens voids (Figure S16-17) as it terminates rather than bridges bonding networks, reducing $k$ to 0.51 W/mK at 20%. Oxygen, in contrast, reduces $k$ monotonically to 0.68 W/mK at 20% with no improvement at any concentration. Boron bonds preferentially with oxygen over nitrogen, disrupting the B-N network even at low concentrations. The resulting B-O bonds reduce $sp^3$ content and eliminate three-dimensional heat transport pathways, forcing nitrogen into undercoordinated configurations and lowering density through bond length mismatch and void formation.

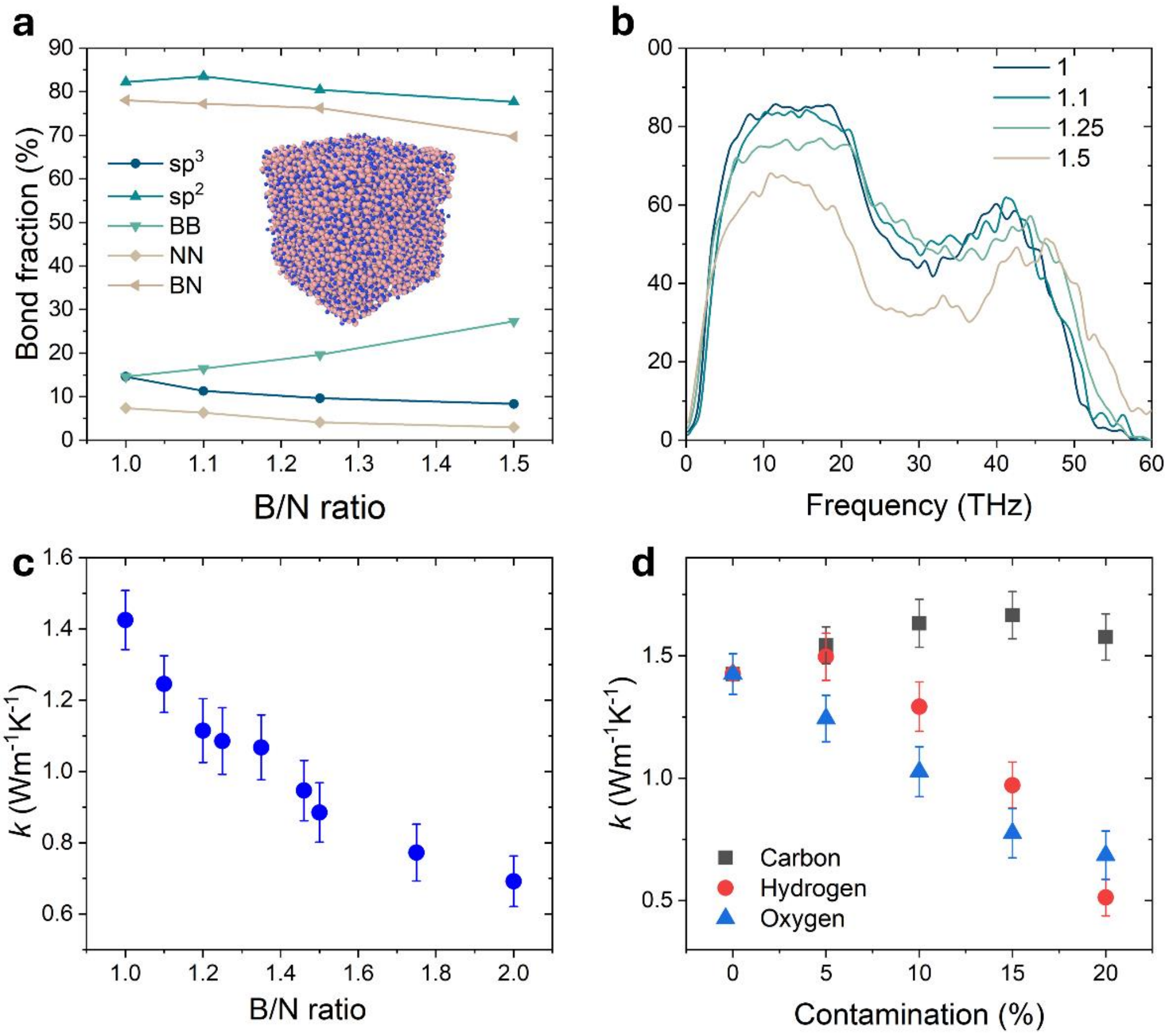

**Figure 5. Molecular Dynamics simulations.** (a) Change in bonding environment and atomic hybridization according to stoichiometry of a-BN Inset: atomic structure of stoichiometric a-BN. (b) Vibrational density of states as a function of B/N ratio. (c) Thermal conductivity as a function of ratio between boron and nitrogen. (d) Thermal conductivity as a function of concentration of contaminant atoms.

Overall, the MD results are in good quantitative agreement with the FDTR measurements for the a-BN samples. The predicted decrease of density and nanovoid formation resulting from non-stoichiometric BN and small presence of contaminants also helps to understand the relatively low BLS values of Young modulus obtained for a-BN samples. Interestingly, the MD simulations provide a route for improved thermal performance in this material, which would rely on minimizing the homonuclear B-B and N-N bonds, either by maintaining the strict B/N ratio or by incorporating carbon atoms in the system.

In summary, we demonstrate how the synthesis temperature drives BN from a fully amorphous phase to a mixed amorphous–crystalline state, with direct consequences for cross-plane heat transport and mechanical stiffness. Across this transition, BN films exhibit improvement both in cross-plane thermal conductivity and elastic modulus with increasing structural order. Notably, the cross-plane thermal conductivity of low-temperature-grown a-BN remains relatively close to that of thin h-BN and mixed-phase BN, rather than decreasing by orders of magnitude. This limited thermal penalty is particularly important for integration, because a-BN can be synthesized at substantially lower temperatures than h-BN. Green–Kubo simulations further show that these trends originate from the evolution of bonding topology and vibrational transport, while identifying stoichiometry and oxygen, hydrogen, and carbon incorporation as practical parameters to tune functionality. Together, these results position a-BN and mixed-phase BN as technologically relevant dielectric and coating platforms.

**Acknowledgments**

The Catalan Institute of Nanoscience and Nanotechnology (ICN2) is funded by the CERCA program/Generalitat de Catalunya and is supported by the Severo Ochoa Centres of Excellence Programme, Grant CEX2021-001214-S, funded by MCIU/AEI/10.13039.501100011033. This work was supported by the MICIU project ELEMENTAL (Grant No. PID2023-152783OB-I00) and European Commission FLAG-ERA project MINERVA (AEI Grant No. PCI2021-122092-2A and ANR-21-GRF1-0002). OK is supported by the REDI Program, a project that has received funding from the European Union's Horizon 2020 research and innovation programme under the Marie Skłodowska-Curie grant agreement No. 101034328. ECA acknowledges support from RYC2024-048225-I from MICIU/AEI/10.13039/501100011033 and FSE+. MS, BG and SR acknowledge financial support from the project PETIE (National Center for Research and Development, Grant No. EIG CONCERT-JAPAN/9/91/PETITE/2023 and PCI2023-143399 and European Union). PX acknowledges support from the European Union's Horizon Europe research and innovation programme under the Marie Skłodowska-Curie grant agreement No. 101107155 (Carmen). We thank dr Jakub Szewczyk and Adam Krysztofik for the AFM measurements.